\documentclass[12pt]{article}
\usepackage{amssymb}

 \def\be{\begin{equation}}
 \def\ee{\end{equation}}
 \def\bes{\begin{eqnarray}}
 \def\ees{\end{eqnarray}}
  \def\a{\alpha}
  \def\b{\beta}

 \def\G{\Gamma}

 \def\p{\partial}
 
 \def\t{\tau}

 \def\2{\frac{1}{2}}
 \def\4{\frac{1}{4}}

\catcode`\@=11
\def\@citex[#1]#2{%
\if@filesw \immediate \write \@auxout {\string \citation {#2}}\fi
\@tempcntb\m@ne \let\@h@ld\relax \def\@citea{}%
\@cite{%
  \@for \@citeb:=#2\do {%
    \@ifundefined {b@\@citeb}%
      {\@h@ld\@citea\@tempcntb\m@ne{\bf ?}%
      \@warning {Citation `\@citeb ' on page \thepage \space
undefined}}%
      {\@tempcnta\@tempcntb \advance\@tempcnta\@ne%
      \@tempcntb\number\csname b@\@citeb \endcsname \relax%
      \ifnum\@tempcnta=\@tempcntb 
it
        \ifx\@h@ld\relax%
          \edef \@h@ld{\@citea\csname b@\@citeb\endcsname}%
        \else%
          \edef\@h@ld{\ifmmode{-}\else--\fi\csname
b@\@citeb\endcsname}%
        \fi%
      \else
        \@h@ld\@citea\csname b@\@citeb \endcsname%
        \let\@h@ld\relax%
      \fi}%
    \def\@citea{,\penalty\@highpenalty\,}%
  }\@h@ld
}{#1}}

\def\@citeb#1#2{{[#1]\if@tempswa , #2\fi}}
\def\@citeu#1#2{{$^{#1}$\if@tempswa , #2\fi }}
\def\@citep#1#2{{#1\if@tempswa , #2\fi}}

\begin{document}
\begin{titlepage}
\begin{flushright}
UTHET-06-0601\\
hep-th/0607139-v2\\
June 2007
\end{flushright}

\vspace{0.20in}
\begin{centering}

{\Large {\bf AdS/CFT Correspondence with Heat Conduction}}
\\

\vspace{0.7in} {\bf James Alsup,$^*$\,\,\,\, George Siopsis$^\flat$}

\vspace{0.1in}

{\em Department of Physics and Astronomy, The University of
Tennessee, Knoxville,\\ TN 37996 - 1200, USA.}

\vspace{0.2in} {\bf and Chad Middleton$^\dag$}

\vspace{0.1in}

{\em Department of Physics, Rhodes College, Memphis, TN 38112-1690, USA.}

\vspace{0.04in}

\vspace{0.04in}

\end{centering}

\vspace{0.4in}

\begin{abstract}

We study an extension of the gravity dual to a perfect fluid model
found by
Janik and Peschanski.
By relaxing one of the constraints, namely invariance under reflection in the longitudinal direction,
we introduce a metric ansatz which includes off-diagonal terms.
We also include an $R$-charge following Bak and Janik.
We solve the Maxwell-Einstein equations and
through holographic renormalization, we show that the off-diagonal components of the bulk metric give rise to heat conduction in the corresponding CFT on the boundary.

 \end{abstract}

\vfill\footnoterule {\footnotesize
%
%
$^{*}$~jalsup1@utk.edu\\
$^\flat$~siopsis@tennessee.edu\\
$^\dag$~MiddletonC@rhodes.edu

}
\end{titlepage}

\section{Introduction}

It is intriguing that strongly interacting particles produced in
heavy ion collisions exhibit perfect fluid
characteristics which are appropriately described by a hydrodynamic model.
The hydrodynamic behavior in the context of
Quantum Chromodynamics (QCD) was studied by Bjorken in the early eighties \cite{Bjorken}.
It remains a challenge to derive such behavior from first principles in QCD.
On the other hand, recent experimental results at RHIC have provided experimental evidence of a hydrodynamic form of strongly interacting matter~\cite{hydro}.

Much has been learned about gauge theories in the strong coupling regime through
the AdS/CFT correspondence~\cite{adscft}\cite{adscftrev}
which enables us to understand these gauge theories by studying their gravity duals
in AdS space with one additional dimension.
Strictly speaking, this correspondence applies to the maximally symmetric $\mathcal{N}=4$ super Yang-Mills theory. Nevertheless, its application to gauge theories with less symmetry, including QCD, has been considered.
In particular, regarding heavy ion collisions, a gravity dual has been used to
extract information about jet quenching \cite{sinzahed}, transport coefficients~\cite{sonstarinets}, the fireball produced at RHIC~\cite{zahedsin}, as well as a bound on $\eta/s$ \cite{son}.

In an interesting work, Janik and Peschanski \cite{JP} discussed a solution to the Einstein equations in the bulk AdS space
in the limit $\tau\to\infty$,
where $\tau$ is the proper time in the longitudinal plane.
By demanding regularity in the bulk, they showed that the acceptable solution
corresponds through holographic renormalization~\cite{Skenderis} to a perfect fluid on the boundary of AdS.  The work was furthered by Bak and Janik \cite{BJ} who studied the Maxwell-Einstein equations in the bulk with a conserved R-charge.

Here we extend the results of~\cite{JP}\cite{BJ} by relaxing the constraint of invariance under reflection in the longitudinal direction.
This allows us to include off-diagonal terms in the bulk metric ansatz.
We solve the Maxwell-Einstein equations in the bulk in the limit $\tau\to\infty$.
We obtain an exact solution and show, through holographic
renormalization, that it corresponds to a non-viscous fluid of non-vanishing chemical potential.
The novel characteristic is a temperature gradient in the longitudinal direction.
We study the thermodynamic properties of the fluid and their relation to
the form of the bulk metric.

\section{R-Charged Perfect Fluid}
\label{sec2}

We are interested in understanding the behavior of a fluid described by a gauge theory in
a
four-dimensional space spanned by coordinates $x^\mu$,
($\mu = 0,1,2,3$).
Following~\cite{Bjorken}, we introduce the proper time $\tau$ and rapidity $y$
on the longitudinal plane, defined by
\be\label{eqtauy}
x^0=\t \cosh y ~~~~~ , ~~~~~x^1=\t \sinh y
\ee
The transverse coordinates will be denoted by $x^\bot = (x^2,x^3)$.
The gravity dual of the four-dimensional theory will be five-dimensional.
Let $z$ denote the fifth dimension.
We shall solve the Maxwell-Einstein equations in the bulk
\be\label{Einstein} R_{\a\b}+4g_{\a\b}+\frac{1}{12}F_{\mu\nu}F^{\mu\nu}g_{\a\b}-\frac{1}{2}F_\a^\mu F_{\b\mu}=0
\ee
\be\label{Maxwell}
\nabla_\a F^{\a\mu}-\frac{\sqrt 3}{48\sqrt{-g}}\epsilon^{\mu\a\b\gamma\delta}F_{\a\b}F_{\gamma\delta}=0
\ee
Following~\cite{BJ}, we adopt the metric and field strength tensor ansatz
\bes\label{metric}
ds^2&=&\frac{1}{z^2}\left[-e^{a(\tau,z)}d\tau^2+\tau^2e^{b(\tau,z)}dy^2+dx^{\bot 2}+e^{d(\tau,z)}dz^2\right]\\
F_{z\tau}&=&K(\tau,z)
\ees
which is the most general bulk metric obeying boost invariance,
symmetry under reflection in the longitudinal direction ($y\rightarrow -y$), plus translational and rotational
invariance.
It is convenient to introduce the coordinate
\be v=\frac{z}{\tau^\eta}\ee
in terms of which the metric~(\ref{metric}) reads
\bes\label{metric1}
ds^2&=&\frac{1}{v^2\tau^{2\eta}}\left[-(e^{a(\tau,v)}-e^{d(\tau,v)}\eta^2v^2\tau^{2(\eta-1)})d\tau^2+\tau^2e^{b(\tau,v)}dy^2
+dx^{\bot 2}+e^{d(\tau,v)}\tau^{2\eta}dv^2\right]\nonumber\\
&+&2\eta e^{d(\tau,v)}\frac{dvd\tau}{v\tau} \\
F_{v\tau}&=&F(\tau,v)
\ees
Substituting this ansatz into the Maxwell-Einstein
equations,
we obtain the
leading behavior in the $\tau\rightarrow\infty$
limit~\cite{JP} \cite{BJ} \bes\label{eqperfl}
\eta=1/3,~~&~&~~F(\tau,v)=\frac{q v}{\tau^{1/3}}\nonumber\\
a(v)=\ln\left(1+\a v^4+\frac{q^2}{12}v^6\right),~~~b(v)&=&0,~~~
d(v)=-\ln\left(1+\a v^4+\frac{q^2}{12}v^6\right)
\ees
in terms of arbitrary parameters $\alpha$ and $q$.

The above bulk metric may be related to the vacuum expectation value of the stress-energy tensor of the corresponding gauge theory on the boundary
through holographic renormalization~\cite{Skenderis}.
This is done as follows. The metric~(\ref{metric}) is transformed to that of the general asymptotically AdS metric in Fefferman-Graham coordinates
\be\label{FG}
ds^2=\frac{g_{\mu\nu}dx^\mu dx^\nu+dz^2}{z^2}\ee
Near the
boundary at $z=0$ we may expand \be
g_{\mu\nu}=g_{\mu\nu}^{(0)}+z^2g_{\mu\nu}^{(2)}+z^4g_{\mu\nu}^{(4)}+\dots \ee
where $g^{(0)}_{\mu\nu}=\eta_{\mu\nu}$ and $g^{(2)}_{\mu\nu}=0$.
The next coefficient, $g_{\mu\nu}^{(4)}$, is proportional to the vacuum expectation value of the stress-energy tensor. Thus,
\be\label{eqgT} \langle T_{\mu\nu}\rangle =
\frac{N_c^2}{2\pi^2}g_{\mu\nu}^{(4)}
\ee
It is can be seen from the form of $g_{\mu\nu}^{(4)}$, for the special value~(\ref{eqperfl}) of $\eta$,
that the stress-energy tensor corresponds to that of a perfect fluid
\be\label{eqTperfl}
T_{\a\b}=(\varepsilon+p)u_\a u_\b+p \eta_{\a\b}
\ee
obeying the equation of state $p= \frac{1}{3} \varepsilon$ (tracelessness due to conformal invariance).
Further, from the explicit form of the metric it follows that the energy density and temperature fall off, respectively, as
\bes\label{eqtaurhoT} \varepsilon =\frac{3N_c^2}{8\pi^2v_h^4}\left( 1+\frac{q^2 v_h^6}{12}\right) \ \tau^{-4/3} 
\\
\label{eqT}T=T_H = \frac{1}{\pi v_h} \left( 1-\frac{q^2 v_h^6}{24}\right) \ \tau^{-1/3} \ees
where $v_h$ is the position of the horizon (obeying $1+\a v_h^4+q^2 v_h^6/12 =0$).

After performing a detailed thermodynamic analysis one finds the entropy density ($s$), charge density ($\rho$), and chemical potential ($\mu$), respectively,
\bes\label{eqtausqu}
s=\frac{S_{bulk}}{V_{brane}}=\frac{N_c^2}{2\pi v_h^3\tau},~~~\rho=\frac{D^{z\tau}}{V_{brane}}=\frac{N_c^2 q}{8\pi^2\tau},~~~\mu=A_\tau(z_h)-A_\tau(z=0)=\frac{q v_h^2}{2\tau^{1/3}}
\ees
Expressing the energy density in terms of $s$ and $\rho$,
\be\label{murho}
\varepsilon(s,\rho)=\frac{3s^{4/3}}{2(2\pi N_c)^{2/3}}\left(1+\frac{4\pi^2\rho^2}{3s^2}\right)
\ee
one verifies that the chemical potential is conjugate to $\rho$ ($\mu = \partial\varepsilon / \partial\rho$).

We must take note that that the geometry is evolving and exact notions of temperature and entropy are ill defined.  However, it is assumed that the approximate notions still prevail.  A discussion of dynamical horizons is given in \cite{AK}.

\section{An extension of the bulk metric}

Next, we consider an extension of the bulk metric~(\ref{metric}) which yielded the perfect fluid model discussed above.
To this end, we relax one of the conditions which led to the general form of the metric,
namely the requirement of invariance under reflection in the longitudinal direction, $y\to -y$.
We may then add appropriate off-diagonal terms in the bulk metric~(\ref{metric1})
that will lead to new non-singular solutions to the Maxwell-Einstein equations~(\ref{Einstein}).
We therefore consider the metric and gauge field ansatz 
\bes\label{metric2}
ds^2&=&\frac{1}{v^2\tau^{2\eta}}\left[-(e^{a(v)}-e^{d(v)}\eta^2v^2\tau^{2(\eta-1)})d\tau^2+\tau^2e^{b(v)}dy^2
+dx^{\bot 2}+e^{d(v)}\tau^{2\eta}dv^2\right]\nonumber\\
&+&2\eta e^{d(v)}\frac{dvd\tau}{v\tau} + 2v^{-2} \tau^{\lambda-2\eta}h_{||}(v)d\tau dy\nonumber\\
A_\tau&=&\tau^{\xi_0} A_0(v)\nonumber\\ A_y&=&\tau^{\xi_1} A_1(v)
\ees
where the second line of the metric consists of the off-diagonal terms added to the ``perfect fluid'' ansatz (\ref{metric1}).
The exponents $-1<\lambda-2\eta, \xi_0, \xi_1 \leq0$ of $\tau$, where bounds are placed so that the associated functions do not singularly dominate, and the functions, $a(v),~b(v),~d(v)$ and longitudinal coupling $h_{||} (v)$ are to be determined by the Maxwell-Einstein equations~(\ref{Einstein}).

Substituting the modified ansatz~(\ref{metric2}) into (\ref{Einstein}), we obtain coupled differential equations for the various functions parametrizing the ansatz.
The task of extracting the leading contribution is complicated by the fact that we now have four parameters determining the order of the expansion, $\eta$ (which is present in the perfect fluid case) as well as $\lambda$, $\xi_0$, and $\xi_1$.  For consistency of the expansion, we need
\be
0<\eta<2/5
\ee
to be compared with the acceptable range $0<\eta<1$ in the perfect fluid case.  The restricted range still includes the special value (\ref{eqperfl}).

Let us first consider the $v$ and $\tau$ components of the Maxwell equations.  They may be factored in powers of $\tau$ as
\bes
e^{a(v)+2b(v)}v^4A_0'(v)\left(\xi_0+1-2\eta\right)+\mathcal{O}(\tau^{2(\lambda-1)},\tau^{\lambda+\xi_1-2-\xi_0},\tau^{3\lambda+\xi_1-4-\xi_0})&=&0\nonumber\\
e^{a(v)+2b(v)}v^3\left[A_0'(v)(2+va'(v)-vb'(v)+vd'(v))-2vA_0''(v)\right] && \nonumber\\
+\mathcal{O}(\tau^{2(\lambda-1)},\tau^{\lambda+\xi_1-2-\xi_0},\tau^{3\lambda+\xi_1-4-\xi_0})&=&0
\ees
The $v$ component is satisfied by the choice of parameter
\be
\xi_0=2\eta-1
\ee
The $\tau$ component is solved by
\be
A_0'(v)=qv\ e^{\frac{1}{2}\left(a(v)-b(v)+d(v)\right)}
\ee
where $q$ is an arbitrary integration constant.

Turning attention to the $y$ component of the Maxwell equations,
we observe that the leading behavior is not unambiguously determined without
further information on the parameters.
Indeed, in addition to the manifestly $\mathcal{O} (\tau^0)$ term,
there are two other terms in which the exponent of $\tau$ is not necessarily
negative.
The $y$ component reads
\bes\label{eq23}
&&e^{2a(v)+b(v)}v^3\left[A_1'(v)\left(-2+v a'(v)-v b'(v)-v d'(v)\right)+2vA_1''(v)\right]\nonumber\\
&\ &-\tau^{\lambda-\xi_1-1/3}e^{a(v)+b(v)}v^3\nonumber\\
&&\ \ \ \times \left[-2vA_0'(v)h_{||}'(v)+h_{||}(v)\left(A_0'(v)\left(2+v a'(v)+v b'(v)+v d'(v)\right)-2vA_0''(v)\right)\right]\nonumber\\
&\ &-\tau^{3\lambda-\xi_1-7/3}v^3h_{||}^3(v)\left[A_0'(v)(2+vd'(v))-2vA_0''(v)\right]+\mathcal{O}(\tau^{2(\lambda-1)},\tau^{2(\eta-1)})=0
\ees
A similar ambiguity occurs in the
$y\tau$ and $vy$ components of the Einstein equations.
They read, respectively,
\bes\label{eq20} 
\mathbf{c}_1 (v)\tau^{6\eta-2}+\mathbf{c}_2 (v)\tau^{4\eta+\xi_1-\lambda-1}+\mathbf{a}_1 (v) h_{||} (v) + \mathbf{a}_2 (v) h_{||}' (v) + h_{||}'' (v) + \mathcal{O} (\tau^{2(3\eta+\lambda-2)}) &=& 0\nonumber\\
\mathbf{b}_0(v) \tau^{4\eta+\xi_1-\lambda-1} +\mathbf{b}_1 (v) h_{||} (v) + \mathbf{b}_2 (v) h_{||}' (v) + h_{||}'' (v) + \mathcal{O} (\tau^{2(\lambda-1)}) &=& 0 
\ees
where
\bes\mathbf{c}_1 =\frac{v^2 e^{-a}}{3} h_{||}A_0'^2,~~\mathbf{c}_2 =v^2 A_0' A_1' \ , ~~~&~&~~ \mathbf{b}_0 =\frac{v A_0' }{\eta}\left(v\eta A_1'-\xi_1 A_1\right)\nonumber\\
\mathbf{a}_1 =\frac{8(1-e^d)}{v^2}+\frac{d'-a'-b'}{v}+a'b' \ , ~~&~&~~\mathbf{b}_1 =\frac{b'\left(-\eta+1+\lambda\right)}{v\eta}+\frac{b'}{2}\left(a'+b'+d'\right)-b''\nonumber\\
\mathbf{a}_2 =\frac{-3}{v}-\frac{1}{2}(a'+b'+d') \ , ~~&~&~~\mathbf{b}_2 =\frac{\eta-1-\lambda}{v\eta}-\frac{1}{2}(a'+b'+d')
\ees
%
For consistency of the two equations~(\ref{eq20}), we must have
$\mathbf{a}_2=\mathbf{b}_2$, which leads to the constraint on the parameters
 \be\label{eqlam}
 \lambda=4\eta-1
 \ee
The condition (\ref{eqlam}) is necessary for consistency but not sufficient.
We need to ensure $\mathbf{a}_1 = \mathbf{b}_1$, as well. To this end, we turn to the diagonal components of the Einstein equations which will determine the parameters $\eta$ and $\xi_1$ as well as the functions $a(v),\ b(v)$ and $d(v)$.   As of now there is enough information to determine $\eta$ by demanding finiteness
of the Maxwell scalar $F^2$.
We have
\be
F^2=F_{\alpha\beta}F^{\alpha\beta}=-2q^2 v^6e^{-b(v)}\tau^{6\eta-2}+\mathcal{O}(\tau^{8\eta+\xi_1-4},\tau^{2(\eta+\xi_1-2)})
\ee
which forces $\eta$ (and therefore $\lambda$ on account of eq.~(\ref{eqlam})) to take the special value
\be\label{etaeq}
\eta= \lambda = 1/3
\ee
The $y$ component of the Maxwell equations may now be used to determine the
remaining parameter $\xi_1$.
Since $\xi_1 \le 0$, the exponents of $\tau$ involving $\xi_1$ will be positive
unless
\be
\xi_1=0
\ee
With this choice, all terms shown explicitly in the $y$ component of the Maxwell equations are $\mathcal{O} (\tau^0)$ (leading).

The rest of the Einstein equations are found to $\mathcal{O}(\tau^{-4/3})$ as
\bes
16(e^d-1)+\frac{4}{3}e^{-a}v^4 A_0'^2+2v\left(4a'+b'-d'\right)+v^2\left(-a'^2-a'b'+a'd'-2a''\right)=0\nonumber\\
16(e^d-1)-\frac{2}{3}e^{-a}v^4 A_0'^2+2v\left(a'+4b'-d'\right)+v^2\left(-a'b'-b'^2+b'd'-2b''\right)=0\nonumber\\
8(e^d-1)-\frac{1}{3}e^{-a}v^4 A_0'^2+v\left(a'+b'-d'\right)=0\nonumber\\
16(e^d-1)-\frac{4}{3}e^{-a}v^4 A_0'^2+2v\left(a'+b'-4d'\right)+v^2\left(-a'^2-b'^2+a'd'+b'd'-2a''-2b''\right)=0\nonumber\\
16(e^d-1)+\frac{4}{3}e^{-a}v^4 A_0'^2+2\left(4a'-b'-d'\right)+v^2\left(-a'^2-a'b'+a'd'-2a''\right)=0
\ees
To leading order, these are identical to the perfect fluid case and are solved by~(\ref{eqperfl}).
The off-diagonal component $h_{||} (v)$ of the metric enters in the next-to-leading order.
Once again, we note that the expansion is consistent for the restricted ranges of $\eta,\xi_0, \xi_1$ and $\lambda$.

Turning back to the equations (\ref{eq23}) and (\ref{eq20}), we observe that for the functions (\ref{eqperfl}), the two Einstein equations coalesce and combined with the $y$ Maxwell equation  $A_1(v)$, $h_{||}(v)$ are uniquely determined at leading order.
The resulting differential equations have regular solutions found after some straightforward algebra as
\be\label{eqhr}
h_{||}(v)=\mathcal{A}\left(v^4+\frac{q^2}{12\a}v^6\right)\ee
and
\be
A_1'(v)=-\frac{q\mathcal{A}}{\a}\ v
\ee
where
$\mathcal{A}$ is an arbitrary constant characterizing the departure from the symmetric state (under reflection in the longitudinal direction) discussed in \cite{BJ}.

\section{Hydrodynamics}

In order to
understand the dynamics of the corresponding gauge theory on the boundary we 
invoke holographic renormalization~\cite{Skenderis}.  To do this we must pass back to the Fefferman-Graham coordinates~(\ref{FG}) and obtain the fourth order term in the expansion.  Following the same procedure as in~\cite{BJ}, we redefine
\be
z=z_{FG}\left(1+\frac{\a z_{FG}^4}{8\tau^{4/3}}+\dots \right)
\ee
which effectively flattens the $z$ coordinate and from eq.~(\ref{eqgT}) we may determine the vacuum expectation value of the
gauge theory stress-energy tensor.  The diagonal components maintain the same form that led to the previous energy density and pressure~(\ref{eqtaurhoT}).  The off-diagonal component of the metric demands the vacuum expectation value of the
gauge theory stress-energy tensor develop an off-diagonal component which behaves as
\be\label{eqTg} \langle
T_{y\tau}\rangle= \frac{N_c^2}{2\pi^2}\frac{\mathcal{A}}{\tau}
\ee

In order to understand how our solution relates to the gauge theory fluid, let us choose a stress-energy tensor which includes an arbitrary energy flux in the longitudinal direction,
start by working with an arbitrary stress tensor with diagonal and $\t y$ components,

\be
T^{\mu\nu} =
\left(\begin{array}{cccc}
T^{\t\t} & T^{\t y} & 0 & 0 \\
T^{\t y} & T^{yy} & 0 & 0\\
0 & 0 & T^{22} & 0\\
0 & 0 & 0 & T^{33}
\end{array}\right) \ .
\ee
Recalling our definition~(\ref{eqtauy}) of coordinates $\t$ and $y$, the metric on the Minkowski space of the fluid reads
\be
ds_4^2=-d\t^2+\t^2dy^2+(dx^\bot)^2
\ee
From the local conservation law
\be
\nabla_\a T^{\a\b}=\p_\a T^{\a\b}+\G^\a_{\a\lambda} T^{\lambda\b}+\G^\b_{\a\lambda} T^{\a\lambda} =0
\ee
and using the Christoffel symbols $\G^y_{y\t}=\frac{1}{\t}=\G^y_{\t y}$ and $\G^\t_{yy}=\t$,
we derive relations between the components of the stress tensor.

Choosing $\b=\t$, we obtain
\be
\p_\t T^{\t\t}+\p_y T^{\t y}+\frac{1}{\t} T^{\t\t}+\t T^{yy}  =0
\ee
Setting $\b=y$, we deduce
\be
\p_\t T^{\t y}+\p_y T^{yy} + \frac{3}{\t} T^{\t y}=0
\ee
One more relation is a consequence of conformal invariance.
Demanding tracelessness, we obtain
\be
-T^{\t\t}+\t^2T^{yy}+T^{22} +T^{33}=0
\ee
In order to match with the expected form of the stress-energy tensor from
holographic renormalization,
we observe that the components of the stress-energy tensor to the order we are
considering should not depend on the rapidity $y$ or the transverse coordinates $x^\bot$.
We may then immediately solve for the energy flux component $T^{\t y}$, obtaining
\be
T^{\t y}=\frac{\mathcal{C}}{\t^3} ~~~, ~~~ T_{\t y}=\frac{-\mathcal{C}}{\t}
\ee
with $\mathcal{C}$ being an arbitrary constant.
This behavior matches the prediction~(\ref{eqTg}) of the gravity dual for $T^{\t y}$.

Solving for the diagonal components $T^{yy},T^{ii}$ ($i=2,3$),
we obtain the stress-energy tensor in terms of $\varepsilon = T^{\t\t}$ (energy density) and the arbitrary constant $\mathcal{C}$ (determining the energy flux),
\be
T^{\mu\nu} =
\left(\begin{array}{cccc}
\varepsilon & \frac{\mathcal{C}}{\t^3} & 0 & 0\\
\frac{\mathcal{C}}{\t^3} & -\frac{1}{\t}\p_\t\varepsilon-\frac{1}{\t^2}\varepsilon & 0 & 0\\
0 & 0 & \varepsilon+\2 \t\p_\t \varepsilon & 0\\
0 & 0 & 0 & \varepsilon+\2 \t\p_\t \varepsilon
\end{array}\right) \ .
\ee
The functional form of the energy density $\varepsilon$ is obtained by demanding isotropy,
$\t^2 T^{yy} = T^{22} = T^{33}$.
We deduce
\be \varepsilon \sim \frac{1}{\t^{4/3}} \ee
matching the perfect fluid behavior~(\ref{eqtaurhoT}).
The equation of state is also unchanged, $p = \frac{1}{3}\varepsilon$.
For the energy flux, we obtain the velocity component
\be u^y \sim \tau \frac{T^{\t y}}{T^{\t\t}} \sim \frac{1}{\t^{2/3}}\ee

Summarizing,
we have obtained the following behavior of the components of the stress-energy tensor,
\bes\label{eqThydro}
T^{\t\t}=\frac{\mathcal{B}}{\t^{4/3}}~~~~~&,&~~~~~T_{\t\t}=\frac{\mathcal{B}}{\t^{4/3}}\nonumber\\
T^{yy}=\frac{\mathcal{B}}{3\t^{10/3}}~~~~~&,&~~~~~T_{yy}=\frac{\mathcal{B}}{3}\t^{2/3}\nonumber\\
T^{ii}=\frac{\mathcal{B}}{3 \t^{4/3}}~~~~~&,&~~~~~T_{ii}=\frac{\mathcal{B}}{3 \t^{4/3}}
\ \ \ \ (i=2,3)\nonumber\\
T^{\t y}=\frac{\mathcal{C}}{\t^3}~~~~&,&~~~~T_{\t y}=\frac{-\mathcal{C}}{\t}
\ees
This behavior exactly matches that expected from the gravity dual.
The constants $\mathcal{B}$ and $\mathcal{C}$ are related to the parameters of the bulk metric by
\be \mathcal{B} = \frac{3N_c^2}{8\pi^2 v_h^4}\ \left( 1 + \frac{q^2v_h^6}{12} \right) \ \ , \ \ \ \ \mathcal{C} = \frac{N_c^2}{2\pi^2}\ \mathcal{A} \ee

To gain further insight into the nature of this fluid,
consider the general case of a dissipative relativistic fluid with stress-energy tensor
and current respectively given by~\cite{LL}
\be\label{eqTghydro} T^{\alpha\beta} = (\rho+p) u^\alpha u^\beta + p\eta^{\alpha\beta} + t^{\alpha\beta} \ \ , \ \ \ \ J^\alpha = \rho u^\alpha + j^\alpha \ee
generalizing the perfect fluid expression (\ref{eqTperfl}). $\rho$ is the particle density (for a perfect fluid, $J^\alpha = \rho u^\alpha$) and the dissipative pieces satisfy
\be u_\alpha t^{\alpha\beta} = u_\alpha j^\alpha = 0 \ee
Standard thermodynamic arguments using the relations
\be\label{eqtherm} \varepsilon+p = Ts+\mu \rho \ \ , \ \ \ \ d\varepsilon = Tds+\mu d\rho \ \ , \ \ \ \ dp = sdT + \rho d\mu \ee
lead to the general expressions
\be t^{\alpha\beta} = -\eta [ \nabla^\alpha u^\beta + u^\alpha u^\gamma \nabla_\gamma u^\beta + (\alpha \leftrightarrow \beta) ] - \left( \zeta - \frac{2}{3} \eta \right) [ \eta^{\alpha\beta} + u^\alpha u^\beta] \nabla_\gamma u^\gamma \ee
\be j^\alpha = -\varkappa ( \partial^\alpha + u^\alpha u^\beta \partial_\beta) \frac{\mu}{T} \ee
in terms of the coefficient of thermal conductivity $\varkappa$ and shear and bulk viscosities, $\eta$ and $\zeta$, respectively.
Assuming no particle transport, $\vec J = \vec 0$,
we deduce
\be \rho\vec u = \varkappa ( \vec\nabla + \vec u u^\beta \partial_\beta) \frac{\mu}{T} \ee
where the vectors are three-dimensional.
This can be solved as an expansion in the derivatives of $\mu /T$,
\be\label{equmu} \vec u = \frac{\varkappa}{\rho} \vec\nabla \frac{\mu}{T} + \dots \ee
This expansion is justified because local velocities are small ($|\vec u| \ll 1$).
To match the behavior (\ref{eqThydro}) of the stress-energy tensor, we assume no viscosity, thus setting
\be \eta = \zeta = 0 \ee
Furthermore, the energy flow ought to be in the longitudinal direction, thus
\be u^\bot = 0 \ee
We may determine the energy density and pressure~(\ref{eqtaurhoT}) directly from the form of the bulk metric.
The results are in agreement with the symmetric state considered in \cite{BJ} (see section \ref{sec2}).
The other thermodynamic quantities (temperature, entropy, charge density and chemical potential) acquire dependence on the longitudinal coordinate (as well as $\tau$) due to the heat flux.
Thus, we obtain corrections to the symmetric case (eqs.~(\ref{eqtaurhoT}) - (\ref{eqtausqu})).
Expanding in the $y$-coordinate, we may write
\be\label{Texp}
T=T_0+y T_1+\dots \ , \ \ \ s = s_0+ys_1+\dots \ , \ \ \ \rho=\rho_0+y\rho_1 + \dots \ , \ \ \ \mu = \mu_0 + y\mu_1 + \dots
\ee 
where the zeroth-order contributions are in agreement with their counterparts in the symmetric case.
Higher-order corrections may be determined
from the thermodynamic relations and the form of the stress-energy tensor.

where $T_0$ is the ideal system's temperature and $T_1$ is the first order correction.  The remaining thermodynamic terms are similarly expanded.

Making use of~(\ref{eqtherm}) we obtain
\be
\mu_1=-T_1\frac{s_0}{\rho_0}  \ \ , \ \ \ \ \  \rho_1=-s_1\frac{T_0}{\mu_0}
\ee
Combining with~(\ref{equmu}) we find
\be
u^y=\frac{\varkappa}{\mu_0\tau^2}\left(\frac{\mu_1}{T_0}-T_1\frac{\mu_0}{T_0^2}\right)+\dots=-\frac{\varkappa}{\rho_0 \tau^2}\left(\frac{s_0}{T_0 \rho_0}+\frac{\mu_0}{T_0^2}\right)T_1+\dots
\ee
From (\ref{eqTghydro}), the energy flux in the longitudinal direction is
\be T^{\tau y} = (\varepsilon + p) u^\tau u^y \ee
To leading order, this is given by
\be T^{\tau y} = \frac{4}{3} \varepsilon u^y +\dots = -\frac{4\varkappa\varepsilon}{3\rho_0 \tau^2}\left(\frac{s_0}{T_0 \rho_0}+\frac{\mu_0}{T_0^2}\right)T_1 + \dots \ee
where we used the equation of state $p = \frac{1}{3} \varepsilon$.

Upon comparison with eqs.~(\ref{eqtaurhoT}) - (\ref{eqtausqu}), (\ref{eqTg}), we deduce the temperature and chemical potential gradients,
\be\label{eqtg2}
T_1=\frac{ q^2 \mathcal{A} N_c^2   }{32\pi^4\varkappa}\frac{v_{h}^6(1-q^2v_{h}^6/24)^2}{(1+q^2v_{h}^6/12)^2}\tau^{-1}\ \ , \ \ \ \ \mu_1=\frac{ q \mathcal{A} N_c^2  }{8\pi^3\varkappa}\frac{v_{h}^3 (1-q^2v_{h}^6/24)^2}{(1+q^2v_{h}^6/12)^2}\tau^{-1}
\ee
the entropy and charge density gradients may be found from the thermodynamic relation~(\ref{murho}),
\bes
s_1&=&\frac{q^2\mathcal{A}N_c^4}{64\pi^4\varkappa}\frac{v_{h}^4 (1-q^2 v_{h}^6/24)^2}{(1+q^2 v_{h}^6/12)^2(1+q^2v_{h}^6/24)}\tau^{-5/3}\nonumber\\
\rho_1&=&-\frac{q \mathcal{A}N_c^4}{32\pi^5\varkappa}\frac{v_{h} (1-q^2 v_{h}^6/24)^3}{(1+q^2 v_{h}^6/12)^2(1+q^2v_{h}^6/24)}\tau^{-5/3}
\ees

%
Finally, the energy flux may be written in terms of the temperature gradient (since the pressure gradient is of higher order) as
\be\label{eqEneF} T^{\tau y} = -\kappa_T \partial_y T \ee
where $\kappa_T$ is the standard definition of thermal conductivity in non-relativistic mechanics.
We obtain~\cite{sonstarinets}
\be \kappa_T = \varkappa \left( \frac{\varepsilon+p}{\tau \rho T } \right)^2 \ee
To leading order, we have
\be \kappa_T = \frac{16\pi^2}{q^2 \tau^2 }\frac{(1+q^2v_{h}^6/12)^2}{v_{h}^6(1-q^2v_{h}^6/24)^2} \varkappa \ee
and using the expression (\ref{eqtg2}) for the temperature gradient, we easily see that the energy flux (\ref{eqEneF}) matches (\ref{eqTg}).


\section{Conclusion}

We solved the Maxwell-Einstein equations in $AdS_5$ for long longitudinal proper time using a metric ansatz which was a generalization of the proposal of ref.~\cite{BJ}.
By relaxing the requirement of invariance under reflection in the longitudinal direction, we were able to add off-diagonal terms to the metric.
We found an explicit solution by
keeping leading terms in the Einstein equations.
Through holographic renormalization, we showed that the bulk metric corresponded to a non-viscous gauge theory fluid
with energy flux in the longitudinal direction.
We studied its thermodynamic properties and calculated the standard coefficient of thermal conductivity and the temperature, chemical potential, entropy, and charge density gradients.

It would be interesting to relax the assumption of rapidity invariance of the bulk metric and consider more general forms of the metric by including dependence on transverse spatial coordinates as well. That would lead to viscous behavior as well as a more complicated energy flow.
  Work in this direction is in progress.

\section*{Acknowledgments}
The work of J.~A.~and G.~S.~was supported in part by the Department of Energy under grant DE-FG05-91ER40627.
C.~M.~was supported in part by a Faculty Professional Travel Grant from Rhodes College.  C.~M.~gratefully acknowledges the hospitality of the Department of Physics and Astronomy at the University of Tennessee where part of this work was done.


\end{document}